\documentclass[aps,twocolumn,pre]{revtex4-1}

\usepackage{graphicx}
\usepackage{amsmath,mathtools,esint}
\usepackage{color,ulem}

\begin{document}

\title{Geometricities of driven transport in presence of reservoir squeezing}
\author{Javed Akhtar}
\author{Jimli Goswami}

\author{Himangshu Prabal Goswami}
\email{hpg@gauhati.ac.in}
\affiliation{$^a$Department of Chemistry, Gauhati University, Jalukbari, Guwahati-781014, Assam, India}
\date{\today}

\begin{abstract} 
In a bare site coupled to two reservoirs, we explore the statistics of boson exchange in the presence of two simultaneous processes: squeezing the two reservoirs and driving the two reservoirs. The squeezing parameters compete with the geometric phaselike effect or geometricity to alter the nature of the steadystate flux and noise. The even (odd) geometric cumulants and the total minimum entropy are found to be symmetric (antisymmetric) with respect to exchanging the left and right squeezing parameters. Upon increasing the strength of the squeezing parameters, loss of  geometricity is observed. Under maximum squeezing, one can recover a standard steadystate fluctuation theorem even in the presence of phase different driving protocol. A recently proposed modified geometric thermodynamic uncertainty principle is found to be robust.
\end{abstract}

\maketitle





\section{Introduction}
Phase-different multiparametric temporal driving allows an additional leverage over a system's dynamics \cite{pancharatnam1956generalized,berry1984quantal}. This leverage is due to gauge-invariant geometric observables during the system's time evolution which affect the driven transport and time-dependent energy conversion processes. Additional phases during the time evolution of a system that arise during cyclic variations of two parameter adiabatic driving are usually referred to as Pancharatnam-Berry phases which introduce nontriviality into a well understood system \cite{mukunda1993quantum}.   As a first application,
the holonomy of the parametric space was engineered to observe bias-independent electronic pumping under slow periodic variations \cite{thouless1983quantization}.   
Subsequently, this paradigm was extended to hem in upon nonequilibrium systems that exchange matter/energy with macroscopic reservoirs \cite{carollo2003geometric,wang2022geometric}. Usually, geometric contributions in nonequilibrium quantum systems are introduced by either driving the reservoirs' temperatures or chemical potentials or even the system-reservoir couplings \cite{wang2022geometric}. In such systems, the geometric effects not only actuate the steadystate dynamics but also lead to violations of the well established fluctuation theorems (FT) and thermodynamic uncertainity relationships (TUR) \cite{PhysRevLett.104.170601,PhysRevE.106.024131,hino2020fluctuation}, which are otherwise  robust even in the presence quantum coherences, entanglement and quantum squeezing. These geometric effects are almost entirely quantified by identifying its contribution to the generating function describing any exchange processes in a nonequilibrium quantum system \cite{wang2022geometric,takahashi2020full,hino2020fluctuation}. The resulting generating functions, derived from a full counting statistical (FCS) method, 
have an additive term apart from the inherent dynamic term which is driving dependent and possesses a  geometric curvature in the parameter space \cite{PhysRevLett.104.170601,sinitsyn2007berry}. The geometric contribution in such nonequilibrium quantum systems can also be observed during the evolution of the system's density matrix \cite{goswami2016geometric}. Although observable, it is no longer a phase factor and hence, is referred to as geometric phaselike effect or simply geometricity. 

 Such effects have also been explored in quantum heat engines, thermoelectric devices and molecular junctions \cite{hino2021geometrical,PhysRevE.96.052129, giri2019nonequilibrium,eglinton2022geometric, goswami2016geometric, yuge2012geometrical}. Enhancement of engine's constancy, affecting the coherent contribution to flux,  observation of giant Fano factors and fractional quantization of the flux etc have been reported \cite{lu2022geometric,PhysRevE.96.052129,PhysRevLett.104.170601}. On a separate note, in the absence of geometric effects, general observables like flux, higher order fluctuations, constancy, thermodynamic affinities  are also affected when parameters describing the reservoirs are altered, eg. by introducing quantum mechanical squeezing \cite{walls1983squeezed,schnabel2017squeezed,dodonov1991physical, puri1997coherent, dupays2021shortcuts,abebe2021interaction}. Squeezed reservoirs are also known to introduce additional quantum control which have been exploited to observe nontrivial quantum thermodynamics like additional corrective parameters on the classical fluctuation theorem of the Crooks type \cite{holmes2019coherent} or not leading to Jarzynski-Wojcik type of fluctuation theorem \cite{yadalam2022counting}. Squeezed states of the thermal reservoirs have also been exploited to overcome Carnot limit in  heat engines\cite{manzano2016entropy,niedenzu2016operation,agarwalla2017quantum,klaers2017squeezed,newman2017performance} , violate universal maximum power theories \cite{giri2019nonequilibrium,PhysRevE.96.052129,PhysRevE.106.024131} and introduce higher order correlated photon pairs from  MgO:LiNbO$_3$ crystals \cite{ourjoumtsev2011observation,mehmet2010observation}. To  corral a universal understanding of the role of squeezed initial states in FTs and TURs, several possibilities are currently under conceptualization \cite{huang2012effects,hsiang2021fluctuation,talkner2013statistics}. For example, higher order fluctuations during photon transport can be maximized due to mixing between a qubit and squeezed resonators \cite{wang2021nonequilibrium}.
When treated separately, both geometricity (introduced via tuning the reservoirs) and squeezing the reservoirs inherently affect the quantum thermodynamics of nonequilibrium systems separately. Hence, it is a natural question to ask about the quantum thermodynamics of nonequilibrium systems where squeezed reservoirs are subjected to periodic modulations. This paper is the first to address this question. Since, presence of geometricity make even a simple model, eg. a resonant level coupled to two thermal or electronic reservoirs non-trivial \cite{yuge2012geometrical,goswami2016geometric}, we focus on such a system, where the reservoirs are  squeezed.

In this work, we study the effect of squeezing the reservoirs on the statistics of particle exchange when the temperature of the reservoirs are periodically modulated. The geometricity that manifests itself into the quantum statistics is explored in a toy model which is a bare site coupled to two squeezed reservoirs.  Such a model is passably standard and well-studied in quantum transport \cite{giraldi2014coherence, PhysRevLett.104.170601,pekola2021colloquium}.  
 Our work focuses on identifying the competition between squeezing and driving on the fluctuations of boson exchange within a quantum statistical framework. We implement the acknowledged methodology of full counting statistics (FCS)  within a quantum master equation framework \cite{esposito2009nonequilibrium}. Firstly, in Sec. (\ref{model}) we present our model and the general formalism used. In Sec. (\ref{results}), we show our results and analysis after which we conclude in Sec. (\ref{Conclusion}).

\section{Model and Formalism}
\label{model}
\begin{figure}
 \centering
 \includegraphics[width = 0.48 \textwidth]{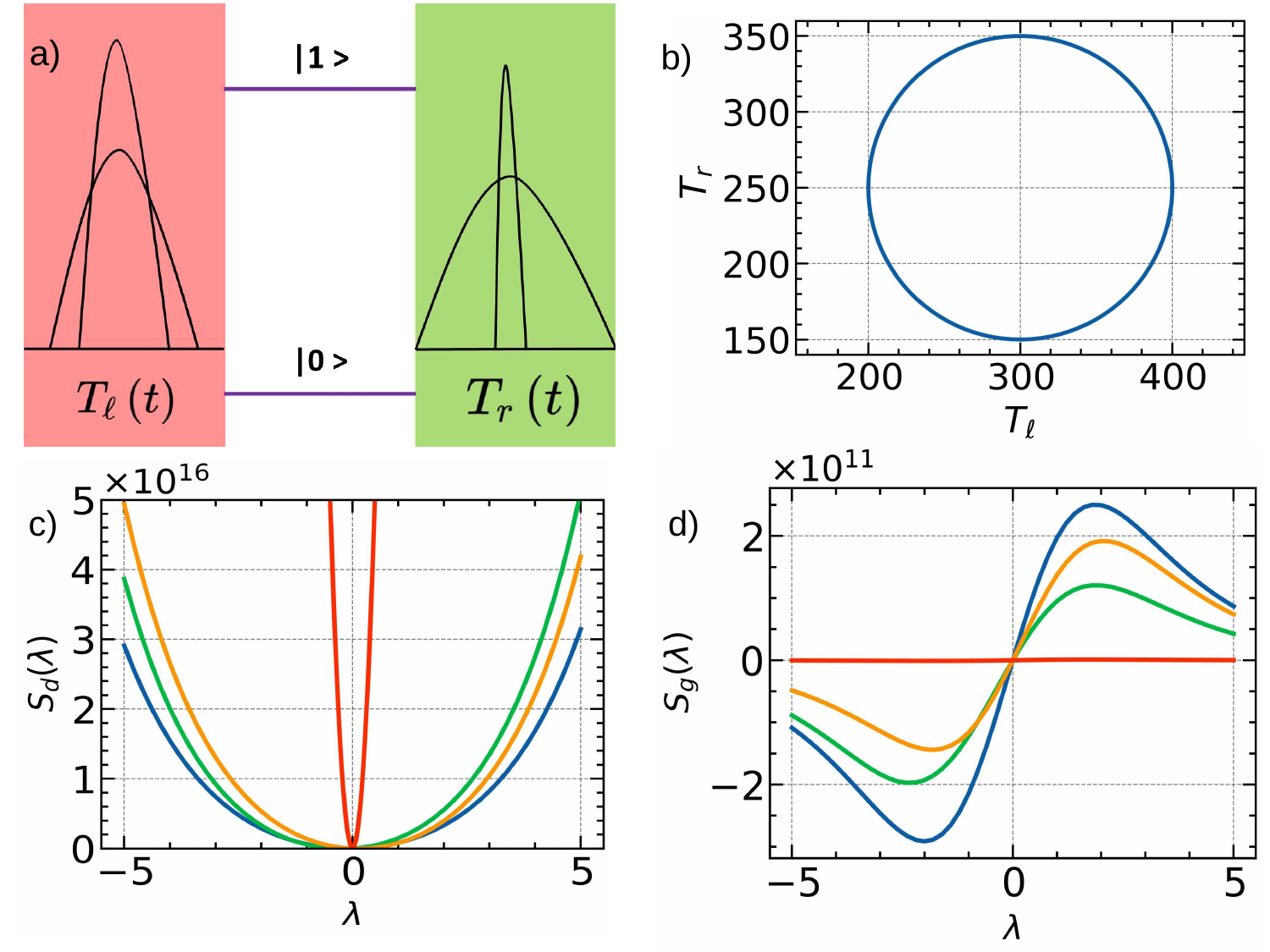}
\caption{ (a) Schematic diagram of two squeezed harmonic baths interacting with a bosonic site with two Fock states ($|0\rangle$ and $|1\rangle$). The temperatures of the two squeezed baths are time-dependent via an amplitude-modulated phase different  driving protocol as per Eq.(\ref{temp_eq})) and (b) represents a circle in the parameter space of $T_\ell$ and $T_r$ with $T_\ell^0=300K, T_r^0=250K$.  Squeezing dependent (c) dynamic (d) geometric cumulant generating function with squeezing parameters ($x_\ell,x_r$) = (0,0),(0.7,0),(0,0.7) and $(\pi,\pi)$ for (c) outermost to innermost curves and (d) in order of decreasing magnetude. The other parameters are fixed throughout the manuscript at $\omega_o=7.4\pi THz$, $\gamma_\ell=\gamma_r=1000 THz$, $\Omega = 100THz$, $A_o=100,\phi=\pi/4$. }
 \label{fig-schematic}
 \end{figure}
A bare site coupled to two reservoirs have been thoroughly studied both in presence and absence of squeezing. \cite{PhysRevE.73.026109,carpio2021quantum,abebe2021interaction,kowalewska2001generalized}. The site can be effectively described by two Fock states that correspond to a boson-occupied ($|1\rangle$) and an unoccupied state, ($|0\rangle$), separated by an energy $\hbar\omega_o$ (see the appendix for the Hamiltonian).  On the experimental front, such a model can be a flexural mode of a GaAs-based nanobeam structure piezoelectrically coupled to squeezed electronic noise (squeezed thermal reservoirs) \cite{klaers2017squeezed} or a qubit system realizable in an NMR setup \cite{saryal2019thermodynamic} as well as in a transmon qubit around a SQUID setup \cite{lu2021nonequilibrium}.  The schematic representation of the model is shown in Fig.(\ref{fig-schematic} a). In such a nonequilibrium system, the time evolution of reduced density matrix, $\hat\rho$, (within standard Born-Markov approximation techniques) is a Pauli rate equation (decoupled from coherences) with two Fock states, $|1\rangle$ and $|0\rangle$ acting as the  boson exchanger between the two squeezed reservoirs. When the reservoirs are driven, the rates become time-dependent (see the appendix).     Within the standard theory of full counting statistical (FCS) formalism \cite{esposito2009nonequilibrium,harbola2007statistics}, one can keep track of the net number of bosons exchanged, $q$, through a moment generating vector for the reduced system in terms of the auxiliary counting field, $\lambda$. In the Liouville space , the reduced moment generating density vector, $|\breve\rho(\lambda,t)\rangle\rangle$, can be written as,
\begin{equation}
\label{lvl-eqn}
    |\dot{\breve\rho}(\lambda,t)\rangle\rangle  =   {\breve{\cal L}} (\lambda)|\breve \rho(0,0)\rangle\rangle,
\end{equation}
where the elements of the density vector contain the populations of the occupied and unoccupied states, $\{\rho_{11}$, $\rho_{00}\}$ (appendix) with the time-dependent evolution superoperator, ${\breve{\cal L}} (\lambda,t)$, given by
\begin{align}
\label{Louv-eq}
{\breve{\cal L}} (\lambda,t)
&=
\left[\begin{array}{cc}
    -\alpha_L(t)-\alpha_R(t) & \beta_L (t)e^{\lambda}+\beta_R(t) \\
    \alpha_L(t)e^{-\lambda}+\alpha_R(t) & -\beta_L(t)-\beta_R(t)
\end{array}\right].
\end{align}
It is a standard practice to ignore the Lamb shifts terms so that the quantum mechanical rates of boson exchange between the system and reservoirs can be recast as:
\begin{align}
\label{eq-alpha}
    \alpha_\nu(t)&=\gamma_\nu\{\cosh{(2x_\nu)(n_\nu^{}(t)+\frac{1}{2})+\frac{1}{2}}\},
\\
\label{eq-beta}
\beta_\nu(t)&=\gamma_\nu \{\cosh{(2x_\nu)(n_\nu^{}(t)+\frac{1}{2})-\frac{1}{2}}\}.
\end{align}
 $\gamma_\nu, \nu =\ell,r$ represents the coupling of the bare site and the $\nu$-th reservoir with $n_\nu=(e^{\hbar\omega_o/T_\nu(t)}-1)^{-1}$
being the Bose-Einstein distribution of the $\nu$-th bath. $x_\nu >0$ is the renormalized parameter responsible for squeezing the $\nu$-th harmonic bath within the Markovian regime \cite{li2017production} (see the appendix). Within this approximation, the squeezing properties get symmetrically distributed about  the concerned left or right squeezed bath's frequency \cite{tanas2002squeezing}. The parametric modulation is present in the reservoirs' temperatures, $T_\nu(t)$, which we take to be of the following form,
\begin{align}
     \label{temp_eq}
     T_{\ell}(t)&:=T_{\ell}^o+A_o\cos(\Omega t+\phi), \\ 
     T_{r}(t)&:=T_{r}^o+A_o\sin(\Omega t + \phi),
\end{align}
$A_{o}$, $\Omega$ and $\phi$ are the amplitude, frequency and phase difference between the driving protocols, respectively. 
Note that, this theory is valid under the adiabatic evolution assumption, where the individual decay timescales of the system and reservoirs are well separated.

In the steady state, when $\lambda= 0$, a zero eigenvalue $\zeta_o(t)$  is obtained from the RHS of Eq.(\ref{Louv-eq}). From this zero-eigenvalue, a cumulant generating function, $S(\lambda)$ within the domain $\lambda\in\{-\infty,\infty\}$, can be constructed  which allows evaluation of the $n$-th order cumulants, $j^{(n)}=\partial_\lambda S(\lambda)|_{\lambda=0}$ \cite{esposito2009nonequilibrium}. In the presence of phase different driving protocol, $S(\lambda)$ is known to be additively separable into two components, one dynamic ($S_d(\lambda,t)$) and a geometric term $S_g(\lambda,t)$.   The geometric term $S_g(\lambda,t)$ essentially is the source of geometricity in such driven dynamics and is obtainable from the left eigenvector ($\langle L_o(\lambda,t)|$) and the right eigenvector ($| R_o(\lambda,t)\rangle$) of the r.h.s of Eq.(\ref{lvl-eqn}) for the eigenvalue $\zeta_o(\lambda,t)$. It is nonexistent if the two parameters (Eq. (\ref{temp_eq}) are driven without any phase difference,i.e $\phi=0$ \cite{yuge2012geometrical}. Both the  dynamic and geometric cumulants can be evaluated as \cite{sinitsyn2007berry,PhysRevLett.104.170601,goswami2016geometric, PhysRevLett.104.170601,hino2020fluctuation, PhysRevE.106.024131,takahashi2020full}
\begin{align}
\label{eq-jd}
 j^{(n)}_d= \partial_\lambda^n S_{d}(\lambda)_{\lambda=0}&=
 \displaystyle\frac{1}{t_p}\int_0^{t_p}\partial_\lambda^n \zeta_o(\lambda,t)|_{\lambda=0}(t) dt\\
 \label{eq-jg}
 j^{(n)}_g= \partial_\lambda^n S_{g}(\lambda)_{\lambda=0}&=
 \displaystyle\frac{1}{t_p}\int_{t_p}^0 \partial_\lambda^n\langle L_o(\lambda,t)| \dot R_o(\lambda,t)\rangle dt|_{\lambda=0}\\
 \label{eq-jg-geo}
 &=
 -\displaystyle\partial_\lambda^n\oiint_{S} {\cal F}_{T_\ell T_r}(\lambda)dT_\ell dT_r|_{\lambda=0}
\end{align}
with $t_p=2\pi/\Omega$ being the time period of the chosen external driving (Eq.(\ref{temp_eq})).  
In Eq. (\ref{eq-jg-geo}), the  integrand, $F_{T_\ell T_r} (\lambda)$, is the known as the geometric curvature  and is analogous to the Pancharatnam-Berry curvature \cite{sinitsyn2007berry,PhysRevLett.104.170601,goswami2016geometric, PhysRevLett.104.170601,hino2020fluctuation, PhysRevE.106.024131,takahashi2020full} in the $T_\ell,T_r$ surface, $S$.
Here, $n=1$ and $2$ correspond to the flux and noise respectively. Both the quantities depend on the  squeezing parameters, $x_\ell$,  $x_r$ through the modified rates in Eq.(\ref{eq-alpha}) and Eq.(\ref{eq-beta}). The dynamic and geometric cumulant generating functions are shown in Fig.(\ref{fig-schematic}c,d) for different squeezing parameters.

{\color{red}} 


\section{Results and Discussion}
\label{results}
\begin{figure}
 \centering
 \includegraphics[width = 0.48 \textwidth]{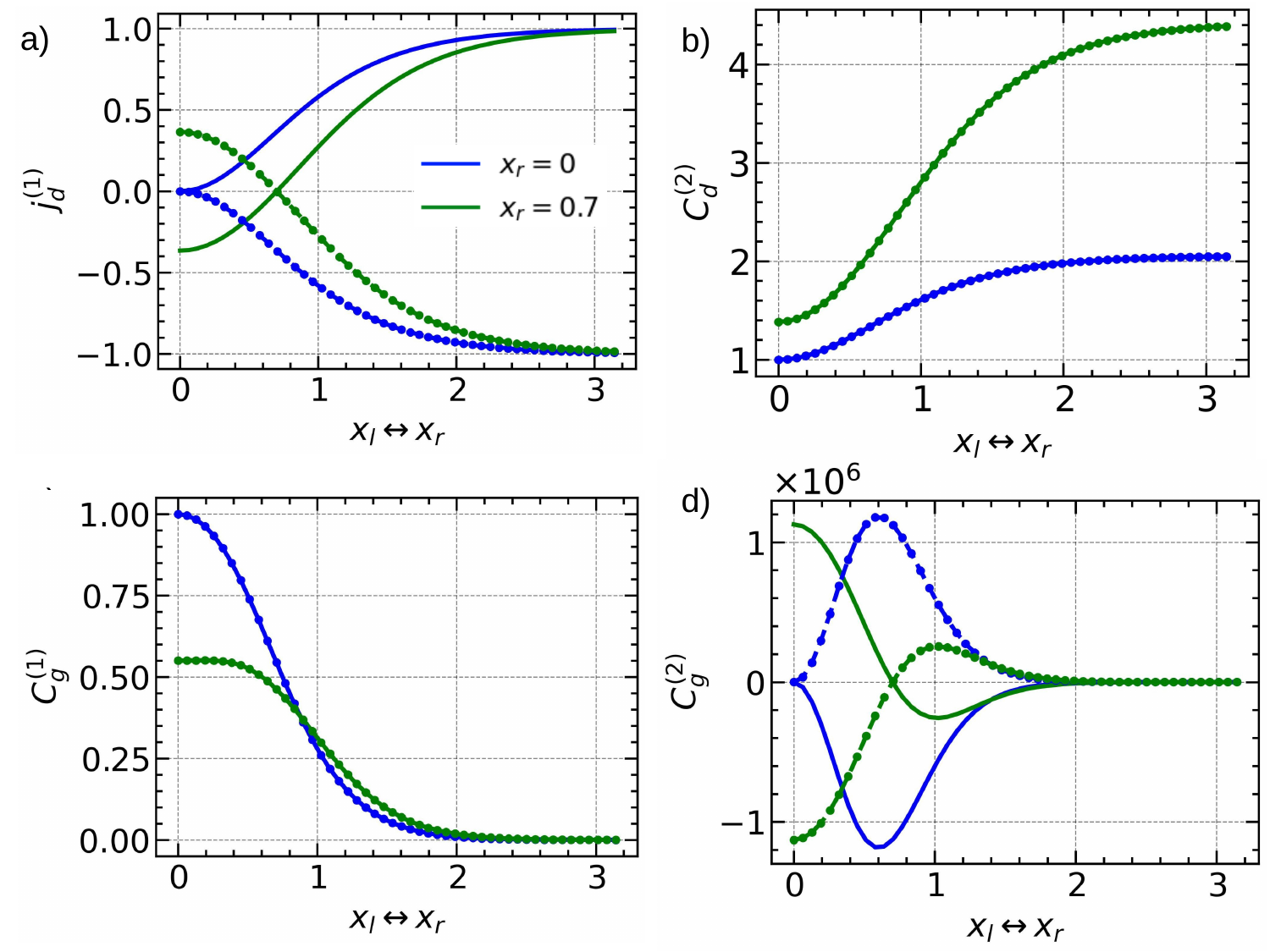}
\caption{ (a) Behavior of the absolute dynamic flux, $j^{(1)}_d$ as a function of the two reservoirs' squeezing parameters  (the $\leftrightarrow$ indicates exchanging the values of the two parameters $x_\ell$ and $x_r$) evaluated at equal initial temperatures $T_\ell^0 =T_r^0=300K$. The solid (dotted) lines are when $x_\ell$ is fixed ($x_\ell$ is changed to $x_r$). Note the antisymmetry due to the exchange $x_\ell\leftrightarrow x_r$. (b) Behavior of the scaled dynamic noise (second cumulant). Note the equality upon exchanging the $x_\ell$ and $x_r$ values. Plot of geometric scaled flux ((c)) and noise ((d)) highlighting the equality and antisymmetry upon exchanging the squeezing parameters. }
 \label{fig-decay}
 \end{figure}

By evaluating the eigensystem of Eq.(\ref{Louv-eq}), we can identify the smallest eigenvalue $\zeta_o(\lambda,t)$ (see the appendix, Eq.(\ref{eq-zeta_o})), from which we numerically obtain the dynamic flux and noise using Eq.(\ref{eq-jd}). The behavior of the two dynamic cumulants ($n=1,2$) are shown in Fig. (\ref{fig-decay}a and b) for equal initial temperatures. The quantitative behavior  is not that different from the undriven case \cite{sarmah2023nonequilibrium} apart from change in magnitude. These also retain the symmetry (antisymmetry) of the even (odd) cumulants with respect to the exchange of the left and right squeezing parameters under equal initial temperature ($T_\ell^0=T_r^0$) setting as well as the saturating behavior as observed earlier for undriven case \cite{sarmah2023nonequilibrium}.  The solid lines in Fig.(\ref{fig-decay}) are evaluated by keeping $x_\ell$ fixed while $x_r$ is varied. The dotted lines represent the case when $x_\ell\to x_r$ while $x_\ell$ is varied, denoted by the symbol $x_\ell\leftrightarrow x_r$ in the abscissa. This is simply because the rates that affect the dynamic cumulants are just scaled by the hyperbolic cosine functions (Eq.\ref{eq-alpha} and Eq.(\ref{eq-beta})) and doesn't alter the overall mathematical structure of the eigenvalue $\zeta_o(\lambda,t)$ (Eq.(\ref{eq-zeta_o})).  In the figures, we also have denoted the cumulants in absence of squeezing ($x_\nu=0$) and driving as $ j^{(n)}_o$ by defining a dimensionless ratio $
 \label{dim-eq}
 C^{(n)}_{d/g}:= j^{(n)}_{d/g}/ j^{(n)}_o$.  When$|C^{(n)}_{d/g}|>(<)1$, the squeezing increases (decreases) the value of the cumulant in comparison to the unsqueezed and undriven case.

\begin{figure}
 \centering
 \includegraphics[width = 0.48 \textwidth]{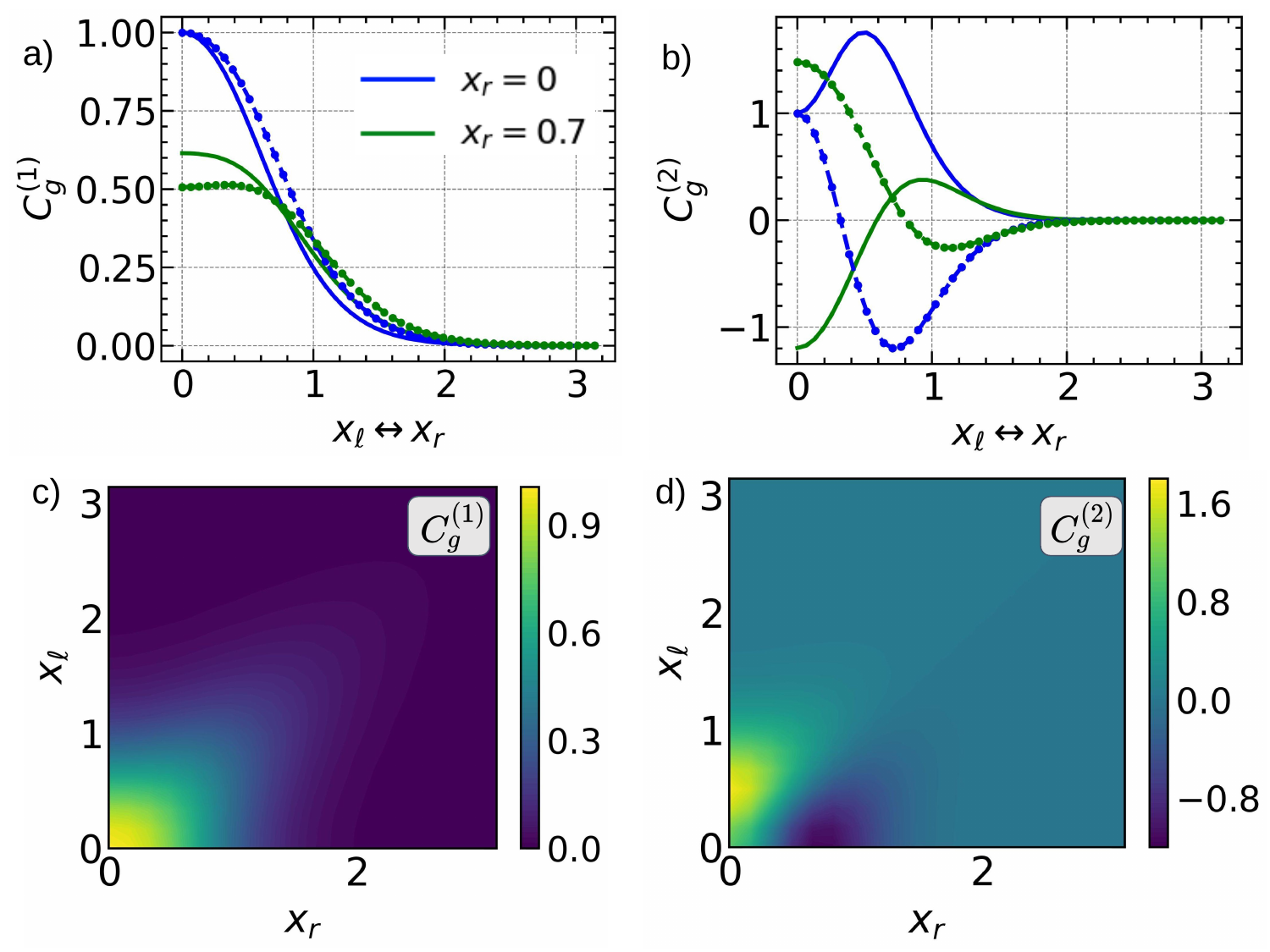}
\caption{ Plot highlighting absence of symmetry and antisymmetry  in the geometric  flux (a) and noise (b) under unequal initial temperatures upon exchanging the squeezing parameters. Contour plots showing the vanishing geometric flux (c) and noise (d) at higher values of squeezing parameters. Note the zero values along the diagonal.}
 \label{fig-con}
 \end{figure}

On substituting the left and right eigenvectors of $\breve {\cal L}(\lambda,t)$ in the geometricity term $\langle L(\lambda,t)|\dot R(\lambda,t)\rangle$, of Eq.(\ref{eq-jg}) we can identify the geometric flux and geometric noise.
The geometric flux is given by,
\begin{align}
\label{eq-jgeo}
 j_g^{(1)}&=-\frac{\Omega}{2\pi}\displaystyle\int_{0}^{t_p}\frac{2\Gamma \cosh (2x_\ell^{}) \cosh (2x_r^{})}{(\gamma_\ell X_\ell^++\gamma_r^{}X_r^+)^3}dt
\end{align}
and shown in Fig. (\ref{fig-decay}c). Note that the geometric flux decays to zero at higher values of the squeezing parameter.  The geometric noise is given by
\begin{align}
\label{eq-ngeo}
 j_g^{(2)}&=-\displaystyle\frac{\Omega}{2\pi}\int_{0}^{t_p}\frac{12 \Gamma^2  \cosh (2x_\ell^{}) \cosh (2x_r^{}) (X_r^+-X_l^+) dt}{(\gamma_\ell+\gamma_r)(\gamma_\ell^{} X_\ell^++\gamma_r^{} X_r^+)^5}
\end{align}
with $\Gamma=\gamma_\ell\gamma_r(\gamma_{\ell}+\gamma_r)$ and $X_\nu^\pm :=\cosh(2x_\nu)(2n_\nu(t)\pm 1)$. The r.h.s of Eq. (\ref{eq-jgeo}) and Eq.(\ref{eq-ngeo}) are evaluated as a function of the squeezing parameters and the scaled function is shown in Fig. (\ref{fig-decay}c) and Fig. (\ref{fig-decay} d) respectively. Both the geometric cumulants are observed decaying to zero as the squeezing parameters are increased. Further, it is also observed that the geometric flux (odd cumulant) is symmetric with respect to exchanging the squeezing parameters while the second cumulant is antisymmetric, contrary to the behavior of the dynamic cumulants. The symmetric behavior upon exchanging $x_\ell$ and $x_r$ in the geometric flux is because the denominator in the integrand inside the r.h.s of Eq.(\ref{eq-jgeo}) is symmetric with respect to exchange. The noise is antisymmetric with respect to exchange because the numerator imparts a negative sign upon exchanging the squeezing parameters. It is interesting to note that both the exchange symmetry and the antisymmetry doesn't hold when the initial temperatures are different. 
This is shown graphically in Fig.(\ref{fig-con}c and d).
 Note that, in Eq.(\ref{eq-ngeo}), when $X_r^+=X_\ell^+$, we obtain $j^{(2)}_g=0$. This condition can be triggered by controlling the squeezing parameters $x_\ell$ and $x_r$  and can be seen as the zero line along the diagonal ($x_\ell=x_r$) of the contour plot in Fig.(\ref{fig-con}d). Under this same condition, the integral in Eq.(\ref{eq-jgeo}) is however nonzero and one observes geometric flux without geometric fluctuations.

We now move on to explain why the geometric effects in the flux and fluctuations vanish at higher squeezing values as seen in Fig.(\ref{fig-decay}c,d) and Fig.(\ref{fig-con}). This is because $S_g(\lambda)$ vanishes at higher values of $x_\ell,x_r$, as seen in Fig. (\ref{fig-schematic}d).  The geometric curvature, in the present model is of the form,
\begin{align}
\label{eq-curvature}
F_{T_\ell T_r}(\lambda)=-\displaystyle\frac{2\Gamma C_\ell C_r \sin(\lambda)}{\{K+4f(\lambda)\}^{3/2}}
\end{align}
with 
\begin{align}
C_\nu&=\frac{1}{k_BT_\nu^2}\hbar\omega e^{\hbar\omega/k_BT_\nu}((n_\nu+1/2)\cosh(2x_\nu)-\frac{1}{2})\\
 K &= \gamma_\nu^{}\displaystyle\sum_{\nu=\ell,r}2\cosh(2x_\nu)(n_\nu+\frac{1}{2})\\
 f(\lambda)&=\displaystyle\prod_{\nu=\ell,r}\gamma_\nu^{}(\cosh(2x_\nu)(n_\nu+1/2)-1/2)\nonumber\\
 &\times e^{\hbar \omega/k_BT_\ell}(e^\lambda-1)+e^{\hbar \omega/k_BT_\ell}(e^{-\lambda}-1)
\end{align}
and is analogous to the known expression for the unsqueezed case ($x_\nu=0$) \cite{sarmah2023nonequilibrium}. $F_{T_\ell T_r} (\lambda)$ is finite for the unsqueezed case around $\lambda =0$. At low values of $\lambda$, the $\sin (\lambda)$ dominates  over the denominator's $f(\lambda)$ term resulting in the typical modified sinusoidal shape as already reported. In the present case too, at lower values of squeezing ($x_\ell,x_r\approx 0$) around $\lambda=0$, such a behavior is shown for $S_g(\lambda)$ as shown in Fig. (\ref{fig-schematic}d). 
As $x_\nu$ is increased, the hyperbolic terms from the squeezing parameters start contributing more to Eq. (\ref{eq-curvature}) around $\lambda=0$ and changes the overall geometricity. These squeezed parameters can now be used to gain control or steer the underlying geometric statistics.

Note that, in general, the mathematical structure of $F_{T_lT_r}$ in Eq.(\ref{eq-curvature}) is such that the numerator (denominator) has an overall squared (cube-halved) dependence on the cosine hyperbolic terms. This structure dictates that, as one keeps squeezing the reservoirs the denominator keeps increasing and hence the amplitude (quantified by the co-efficients) of the $\sin(\lambda)$ term keeps reducing, which results in lower slope around $\lambda=0$. This causes the geometric flux and subsequent cumulants to keep reducing and finally vanishes as shown in Fig.(\ref{fig-schematic}d). We can safely conclude that squeezing the reservoirs reduces the geometricity of the driven system. In this high-squeezing limit, even upon increasing the frequency of phase-different driving, $\Omega\gg 1$, the statistics of exchange is solely governed by the dynamicity of the system, i.e $S_d(\lambda)$. We can prove this analytically by considering the following limiting case. Under the assumption that $n_\nu\ll 1/2$ (low temperature regime), we have,
\begin{align}
 C_\nu|_{n_\nu\ll1/2}^{}&\propto \cosh(2x_\nu)-1 \\
 K|_{n_\nu\ll1/2}^{}&\propto\displaystyle\sum_\nu \cosh(2x_\nu)\\
 f(\lambda)|_{n_\nu\ll1/2}^{}&\propto\displaystyle\prod_{\nu}(\cosh(2x_\nu)-1)
\end{align}
which results in
\begin{align}
\label{eq-curvature-low}
 F_{T_\ell T_r}|_{n_\nu\ll1/2}^{}&\propto\frac{\sin(\lambda)}{\displaystyle\sqrt{\sum_\nu \cosh^3(2x_\nu)}\sqrt{\displaystyle\prod_{\nu}(\cosh(2x_\nu)-1)}}
\end{align}
In the above expression, taking either of the two limits, $x_\ell\to \infty$ or $x_r\to\infty$ results in the r.h.s being zero. Thus, squeezing the reservoirs to its extremum kills the geometric curvature or the geometricity resulting in $S_g(\lambda)=0$.  The complete contour plots of the two geometric cumulants $C^{(1)}_g$ and $C^{(2)}_g$ as a function of $x_\ell$ and $x_r$ are shown in Fig. (\ref{fig-con}c and d). In both the plots, the geometric effects vanish at higher values of squeezing.

\section{Thermodynamic uncertainity relationship}
\label{sec:tur}
For an undriven case, $j^{(n)}_g=0 $ (when $\Omega =0$ or $\phi = 0$), a standard thermodynamic uncertainity relationship (TUR), reminiscent of a steadystate fluctuation theorem holds, given by $F{\cal A}\ge 2k_B $ \cite{pietzonka2017finite,saryal2019thermodynamic} with  $F=j^{(2)}/j^{(1)}$ being the  Fano factor while ${\cal A}$ is the thermodynamic affinity  of the system.
This TUR has been shown not to hold in the presence of geometric effects \cite{PhysRevE.96.052129}. 

In the present case, one can recover the standard TUR in the high squeezing limit of either reservoir.  Under maximum  squeezing, $F_{T_lT_r}(\lambda)=0$ kills the geometric contributions to the system statistics.  We can hence recover a Gallavoti-Cohen symmetry,
\begin{align}
 \label{eq-gc}
 \lim_{x_\nu\to \infty} \displaystyle\frac{1}{t_p}&\int_0^{t_p} \zeta_o(\lambda,t)dt\nonumber\\
 &=\lim_{x_\nu\to \infty}\displaystyle\frac{1}{t_p}\int_0^{t_p}\zeta_o(-\lambda-\displaystyle\lim_{x_\nu\to\infty}{\cal A},t)dt,
\end{align}
with
\begin{align}
 \label{aff-eq}
 {\cal A}=\log \left(\frac{\int_0^{t_p}X_\ell^- X_r^+ dt}{\int_0^{t_p}X_\ell^+ X_r^-dt}\right)
\end{align}
where the time and squeezing dependent quantities $X_\nu^\pm$ are defined in the text below Eq.(\ref{eq-ngeo}).
 Eq.(\ref{aff-eq}) reduces to the known expression $1/T_\ell-1/T_r$ in absence of driving \cite{PhysRevLett.104.170601} that leads to a steadystate fluctuation theorem.  The recovery of the symmetry hence allows us to recover the standard TUR, 
\begin{align}
 \displaystyle\lim_{x_\nu\to\infty}{\cal A} \frac{\displaystyle\lim_{x_\nu\to\infty}j^{(2)}_d}{\displaystyle\lim_{x_\nu\to\infty}j^{(1)}_d}\ge 2k_B
\end{align}

In the case of finite (but not maximal) squeezing, the geometricities are still present. TUR in such a case
 has been shown to get modified by including a geometric correction factor \cite{lu2022geometric}, 
\begin{align}
\label{eq-gen-tur}
 \frac{j^{(2)}\Sigma}{(j^{(1)})^2g(\Omega)}\ge 2k_B
\end{align}
where,
$g(\Omega)$ is the driving dependent geometric correction factor and is of the form,
\begin{align}
\label{eq-geo-tur}
g(\Omega)&=\frac{1}{(1+j^{(1)}_g/j^{(1)}_d)^2}
\end{align}

\begin{figure}
 \centering
  \includegraphics[width = 0.48 \textwidth]{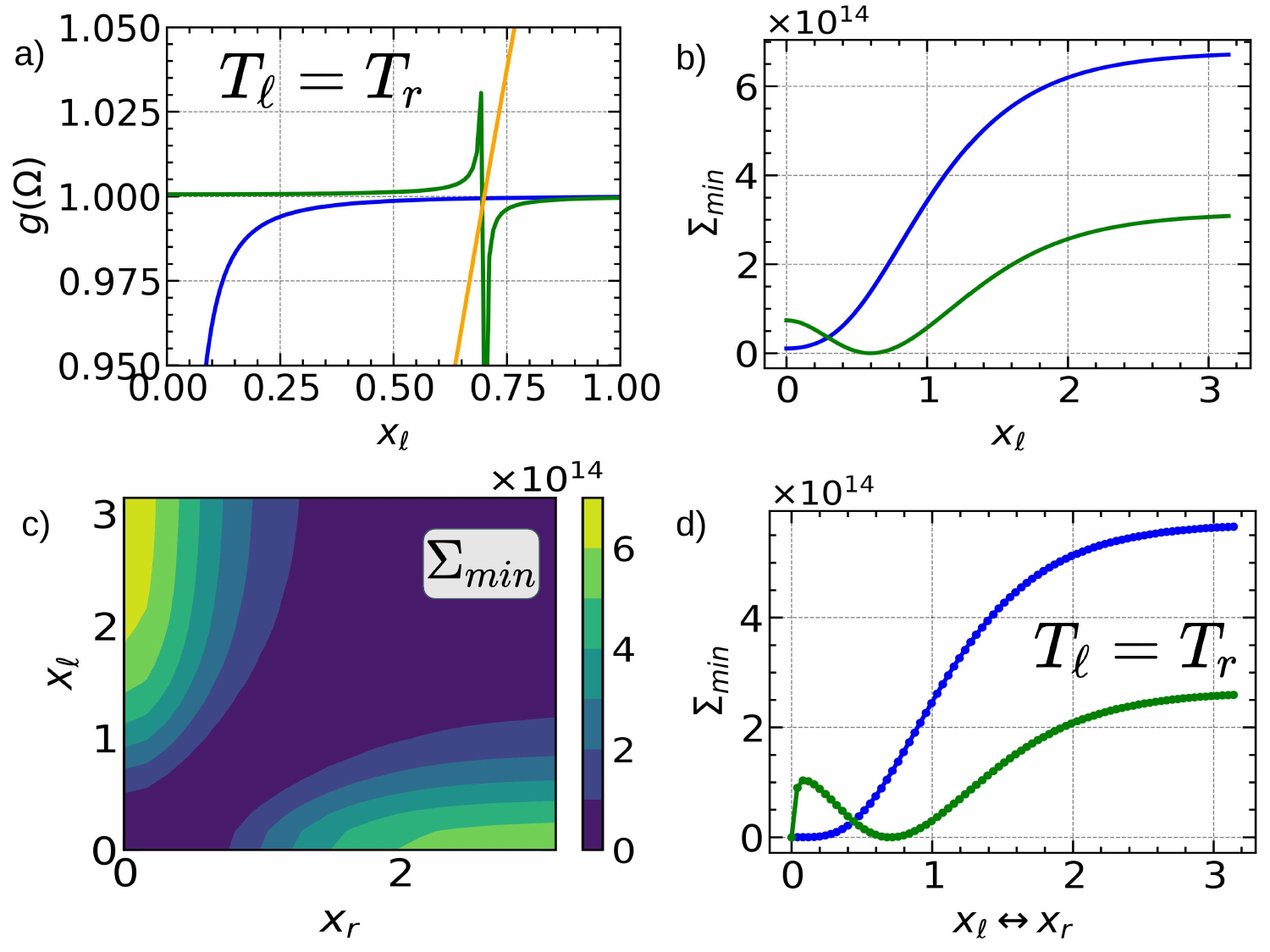}
  \caption{ (a) Behavior of the geometric correction factor, $g(\Omega)$ to the TUR as a function of $x_\ell.$ Note the singularity at $x_\ell=0.7$. At this value of $x_\ell$, the thermodynamic force, ${\cal A} = 0$ (dotted line). (b) Behavior of the minimum entropy produced. It is also zero at $x_\ell =0.7$. (c) Contour of $\Sigma_{min}$ as a function of squeezing parameters. The region where it is zero is where ${\cal A}=0$. (d) Symmetry in the minimum entropy upon exchanging the values of the squeezing parameters.    }
 \label{fig-TUR}
 \end{figure}
We numerically evaluate Eq.(\ref{eq-geo-tur}) and plot $g(\Omega)$ as a function of $x_\ell$ in Fig.(\ref{fig-TUR}a) where a discontinuity is observed at $x_\ell = 0.7$. 
This discontinuity is at that point of $x_\ell$, where ${\cal A} = 0$ ($e^{\cal A}$ is shown as a vertically slanted line) that results in $j^{(1)}_d=0$ in Eq.(\ref{eq-geo-tur}). Further, for any fixed value of $j^{(1)}_g$ in  Eq.(\ref{eq-gen-tur}), 
the r.h.s is greater (less) than unity when $j^{(1)}_d < (>)0$ and vice versa. We have earlier shown that, for an undriven case, the direction of the dynamic flux $j^{(1)}_d$ is controllable through the squeezing parameters due to the modification of the thermodynamic affinity, ${\cal A}$ \cite{sarmah2023nonequilibrium}.  Thus by controlling $x_\ell$, we observe regions where $g(\Omega)>1 $ and $g(\Omega)<1$ characterized by a shift between these two regions at that value of $x_\ell$ where ${\cal A}=0$ as seen in Fig.(\ref{fig-TUR}a). The curve below unity is evaluated by maintaining positive geometric flux and ${\cal A}>0$ ($T_\ell=T_r,x_\ell =0.7,x_r=0$) so that $g(\omega)<1$. In a standard context without squeezing, as in the work of Lu et al \cite{lu2022geometric} where  $g(\Omega)$ was introduced, the dynamic flux is solely dependent on the temperature gradient which controls the thermodynamic affinity and the $g(\Omega)$ is a continuous function either below unity or above unity depending on the signs of the dynamic or geometric fluxes.

We state that this observed discontinuity doesn't lead to  violation of the modified TUR, Eq. (\ref{eq-gen-tur}). Although, not highlighted in the earlier work \cite{lu2022geometric}, the continuity of $g(\Omega)$ in Eq. (\ref{eq-gen-tur}) as a function of a system parameter is rather limited to positive dynamic flux, characterized by ${\cal A}>0$.    As long as we maintain ${\cal A}>0$, by properly choosing $x_\ell$ and $x_r$ values, the modified TUR given by Eq.(\ref{eq-gen-tur}) always holds within these two separate regions. By maintaining, ${\cal A}>0$, we directly estimate the minimum entropy production, $\Sigma_{min}$ in the presence of geometricities, 
\begin{align}
 \label{eq-min-entropy}
 \Sigma_{min}=2k_B\displaystyle\frac{\left(j^{(1)}_d+j^{(1)}_g\right)^2}{j^{(2)}_d+j^{(2)}_g}g(\Omega)
\end{align}
In the above equation, it is not possible to separate the entropy rates into dynamic and geometric contributions. Although when $\Omega\gg 1$, $j^{(1)}_g\gg j^{(1)}_d$, the same cannot be said for the second cumulant which makes the denominator in Eq.(\ref{eq-min-entropy}) to have combined  dynamic and geometric contributions. Nonetheless, the modified TUR allows an easy way to evaluate the total minimum entropy production rate.  Note that, in the presence of geometricities, evaluation of entropies with contribution from both dynamic and geometric components is not at all straightforward \cite{yuge2013geometrical} due to production of excess entropies. We evaluate Eq.(\ref{eq-min-entropy}) and plot it as a function of the left reservoir's squeezing parameters in Fig.(\ref{fig-con})b,c and d). The dependence of $\Sigma_{min}$ on $x_\ell$ is nonlinear and saturates at higher values. In Fig.(\ref{fig-con})c), we show a contour map of $\Sigma_{min}$ for a wide range of $x_\ell$ and $x_r$ values. There exists a wide region of $\Sigma_{min}$ around the diagonal of the contour is where ${\cal A}\approx 0$ that results in $\Sigma_{min}$=0. This region is actually not allowed since $g(\Omega)$ is strictly not defined. One shouldn't substitute this zero value in Eq. (\ref{eq-gen-tur}) and claim it as a violation of the TUR. In Fig.(\ref{fig-con})d), we show the existence of the exchange symmetry ($x_\ell\leftrightarrow x_r$) in the entropy under equal temperature setting. 

\section{Conclusion}
\label{Conclusion}
We employ a full counting statistical method to derive a tilted driven quantum master equation for a simple bosonic site coupled to two squeezed harmonic reservoirs. The temperatures of the two squeezed reservoirs are assumed to be adiabatically driven with a phase-different driving protocol. This allowed us to explore the combined effect of squeezing parameters and the geometricities or geometric phaselike contributions to the steadystate observables, the flux (first cumulant) and the noise (second cumulant). The dynamic cumulants exhibit similar qualitative behavior as a function of squeezing parameters to that of what is already known for an undriven scenario, albeit with modified magnitudes. The geometric cumulants are however affected by the squeezing parameters.  The odd (even) geometric cumulants  are found to be antisymmetric (symmetric) with respect to exchanging the left and right squeezing parameters when the initial thermal gradient is maintained at zero. These also decay to zero as we keep increasing the strength of the reservoirs' squeezing parameters. This is because an increased squeezing prohibits the generation of geometricity in the cumulant generating function. Hence, under maximum squeezing, one can recover a standard steadystate fluctuation theorem which also leads to a standard thermodynamic uncertainty relation even in the presence of phase different driving protocol. Using a recently proposed modified geometric thermodynamic uncertainty principle, which is robust in the presence of squeezing, we estimate the minimum entropy production rate at finite values of dynamic flux. This minimum entropy production rate cannot be separated into dynamic and geometric contributions. It exhibits a saturating behavior and is also symmetric with respect to exchange of the left and right squeezing parameters under a zero initial thermal gradient scenario. 
\section*{Acknowledgments}
   HPG acknowledge the support from the University Grants Commission, New Delhi for the startup research grant, UGC(BSR), Grant No.  F.30-585/2021(BSR) and the Science and Engineering Research Board for the startup grant with file number SERB/SRG/2021/001088.

\appendix
\section{Appendix}

The Hamiltonian of the bare site  interacting with two bosonic reservoirs  can be written as,
\begin{align}
\label{ham-eq}
    \hat{H}&=\displaystyle\hbar\omega_o\hat{b}^\dag\hat{b}+\sum_{i=\nu,\nu\in L,R}^{}\hbar\omega_i^{}\hat{a}_i^\dag\hat{a}_i^{}+\hat{V},
\end{align}
with
\begin{align}
       \label{V-def}
    \hat{V}&=\sum_{i,\nu\in L,R}k_{i}^\nu(\hat{a}_{i\nu}^\dag\hat{b}+\hat{a}_{i\nu}\hat{b}^\dag)
\end{align}
Here, $\hbar\omega_o\hat{b}^\dag\hat{b}$ is the on-site Hamiltonian with bare frequency $\omega_o$, while $\hat{b}^\dag(\hat{b})$ is the bosonic creation (annihilation operator) on the site. The second term is the reservoir Hamiltonian with squeezed harmonic states and is a sum of two terms that represent the left (L) and right (R) squeezed reservoirs.  The single particle operators $\hat{a}_{i\nu}^\dag(\hat{a}_{i\nu})$ represent the creation (annihilation) of a boson in  the i-th mode from (of) the $\nu$-th bath. $\hat{V}$ is the system bath coupling Hamiltonian with  $k_i ^ \nu$ being the coupling constant for the i-th squeezed mode of the $\nu$th bath to the bare site mode.   The squeezed density matrix for the $\nu$-th reservoir ($\hat H_\nu$ being the $\nu$th reservoir Hamiltonian) is given by
\begin{align}
    \hat\rho_\nu&=\frac{1}{Z} \exp\{-\beta_\nu^{} (t) \hat{S}_\nu\hat H_\nu\hat S^\dag_\nu\},\\
    \hat S_\nu &=\displaystyle \prod_{k} e^{\frac{1}{2}(x_{\nu}^* \hat a_{k\nu}^{\dag2}-h.c)}.
\end{align}
$\beta_\nu (t) = (k_B T_\nu(t))^{-1}$ being the inverse temperature and $\hat S_\nu$ is the squeezing operator on the $k-$th mode of the $\nu-$th bath, 
with $x_{\nu}$ being the squeezing the $\nu$-th reservoir's squeezing parameter \cite{dodonov2002nonclassical,li2017production,yadalam2022counting,sarmah2023nonequilibrium}. Assuming that the initial density matrix is factorisable and there is well separation between the system and reservoir timescales (adiabaticity)   \cite{li2017production}, we can write down two adiabatic Pauli-type master equations, with time-dependent squeezed rates, 
\begin{align}
    \dot{\rho}_{11}&=-(\gamma_L(1+N_L(t))+\gamma_R(1+N_R(t)))\rho_{11}\\ \nonumber&+ (\gamma_LN_L(t)+\gamma_RN_R(t))\rho_{00}
\\
    \dot{\rho}_{00}&=(\gamma_L(1+N_L(t))+\gamma_R(1+N_R(t)))\rho_{11}\\ \nonumber&- (\gamma_LN_L(t)+\gamma_RN_R(t))\rho_{00}
\end{align}
where $\langle m |\rho 
|m\rangle =\rho_{mm}$ represents the probability of occupation of the occupied and unoccupied Fock states. Note that the populations and coherences are decoupled and the equations are effectively classical albeit with quantum mechanical rates \cite{bagrets2003full}. 
The driving dependent, squeezed occupation factors are given by \cite{li2017production},
\begin{align}
 \label{eq-bath-cf}
 N_\nu (t)&=\big(\cosh(2x_{i\nu})(n_{\nu}(t)+\frac{1}{2})-\frac{1}{2}\big)
 \end{align}
where $n_\nu(t)$ is the driven Bose function for the $\nu$-th squeezed bath.
Now we can recast  the above two equations in the Liouville space and following the standard procedure of FCS by introducing the auxiliary counting field, $\lambda$ to keep track of the net number of bosons exchanged, $q$ \cite{esposito2009nonequilibrium,harbola2007statistics}, we can arrive at Eq.(\ref{Louv-eq}), where the quantum mechanical rates have been redefined to $\alpha_\nu (t)=\gamma_L(1+N_L(t))$ and $\beta_\nu=\gamma_\nu N_\nu(t)$. 
The $\lambda$-dependent zero eigen value of Eq.(\ref{Louv-eq}) is given by, 

\begin{align}
 \label{eq-zeta_o}
 \zeta_o(\lambda,t)&=-(\gamma_\ell^{} X_\ell^++\gamma_r^{} X_r^+)\\ \nonumber&+\sqrt{(\gamma_\ell^{}+\gamma_r^{})^2+(\gamma_\ell^{} X_\ell^-+\gamma_r^{} X_r^-)f(\lambda)}\\
  X_\nu^\pm &=\cosh(2x_\nu)(2n_\nu(t)\pm 1),\nu=l,r\\
  f(\lambda)&=(\gamma_\ell^{} e^{-\lambda}(1+X_\ell^+)+\gamma_r^{}e^{2\lambda}(X_r^+))
\end{align}
from which the dynamic flux and noise can be numerically evaluated using Eq. (\ref{eq-jd}).


 \bibliographystyle{elsarticle-num} 
 \bibliography{References.bib}

\begin{thebibliography}{10}
\expandafter\ifx\csname url\endcsname\relax
  \def\url#1{\texttt{#1}}\fi
\expandafter\ifx\csname urlprefix\endcsname\relax\def\urlprefix{URL }\fi
\expandafter\ifx\csname href\endcsname\relax
  \def\href#1#2{#2} \def\path#1{#1}\fi

\bibitem{pancharatnam1956generalized}
S.~Pancharatnam, Generalized theory of interference, and its applications: Part
  i. coherent pencils, in: Proceedings of the Indian Academy of
  Sciences-Section A, Vol.~44, Springer India New Delhi, 1956, pp. 247--262.

\bibitem{berry1984quantal}
M.~V. Berry, Quantal phase factors accompanying adiabatic changes, Proceedings
  of the Royal Society of London. A. Mathematical and Physical Sciences
  392~(1802) (1984) 45--57.

\bibitem{mukunda1993quantum}
N.~Mukunda, R.~Simon, Quantum kinematic approach to the geometric phase. i.
  general formalism, Annals of Physics 228~(2) (1993) 205--268.

\bibitem{thouless1983quantization}
D.~Thouless, Quantization of particle transport, Physical Review B 27~(10)
  (1983) 6083.

\bibitem{carollo2003geometric}
A.~Carollo, I.~Fuentes-Guridi, M.~F. Santos, V.~Vedral, Geometric phase in open
  systems, Physical review letters 90~(16) (2003) 160402.

\bibitem{wang2022geometric}
Z.~Wang, L.~Wang, J.~Chen, C.~Wang, J.~Ren, Geometric heat pump: Controlling
  thermal transport with time-dependent modulations, Frontiers of Physics 17
  (2022) 1--14.

\bibitem{PhysRevLett.104.170601}
J.~Ren, P.~H\"anggi, B.~Li,
  \href{https://link.aps.org/doi/10.1103/PhysRevLett.104.170601}{Berry-phase-induced
  heat pumping and its impact on the fluctuation theorem}, Phys. Rev. Lett. 104
  (2010) 170601.
\newblock \href {https://doi.org/10.1103/PhysRevLett.104.170601}
  {\path{doi:10.1103/PhysRevLett.104.170601}}.
\newline\urlprefix\url{https://link.aps.org/doi/10.1103/PhysRevLett.104.170601}

\bibitem{PhysRevE.106.024131}
S.~K. Giri, H.~P. Goswami,
  \href{https://link.aps.org/doi/10.1103/PhysRevE.106.024131}{Controlling
  thermodynamics of a quantum heat engine with modulated amplitude drivings},
  Phys. Rev. E 106 (2022) 024131.
\newblock \href {https://doi.org/10.1103/PhysRevE.106.024131}
  {\path{doi:10.1103/PhysRevE.106.024131}}.
\newline\urlprefix\url{https://link.aps.org/doi/10.1103/PhysRevE.106.024131}

\bibitem{hino2020fluctuation}
Y.~Hino, H.~Hayakawa, Fluctuation relations for adiabatic pumping, Physical
  Review E 102~(1) (2020) 012115.

\bibitem{takahashi2020full}
K.~Takahashi, Y.~Hino, K.~Fujii, H.~Hayakawa, Full counting statistics and
  fluctuation--dissipation relation for periodically driven two-state systems,
  Journal of Statistical Physics 181~(6) (2020) 2206--2224.

\bibitem{sinitsyn2007berry}
N.~Sinitsyn, I.~Nemenman, The berry phase and the pump flux in stochastic
  chemical kinetics, Europhysics Letters 77~(5) (2007) 58001.

\bibitem{goswami2016geometric}
H.~P. Goswami, B.~K. Agarwalla, U.~Harbola, Geometric effects in nonequilibrium
  electron transfer statistics in adiabatically driven quantum junctions,
  Physical Review B 93~(19) (2016) 195441.

\bibitem{hino2021geometrical}
Y.~Hino, H.~Hayakawa, Geometrical formulation of adiabatic pumping as a heat
  engine, Physical Review Research 3~(1) (2021) 013187.

\bibitem{PhysRevE.96.052129}
S.~K. Giri, H.~P. Goswami,
  \href{https://link.aps.org/doi/10.1103/PhysRevE.96.052129}{Geometric
  phaselike effects in a quantum heat engine}, Phys. Rev. E 96 (2017) 052129.
\newblock \href {https://doi.org/10.1103/PhysRevE.96.052129}
  {\path{doi:10.1103/PhysRevE.96.052129}}.
\newline\urlprefix\url{https://link.aps.org/doi/10.1103/PhysRevE.96.052129}

\bibitem{giri2019nonequilibrium}
S.~K. Giri, H.~P. Goswami, Nonequilibrium fluctuations of a driven quantum heat
  engine via machine learning, Physical Review E 99~(2) (2019) 022104.

\bibitem{eglinton2022geometric}
J.~Eglinton, K.~Brandner, Geometric bounds on the power of adiabatic thermal
  machines, Physical Review E 105~(5) (2022) L052102.

\bibitem{yuge2012geometrical}
T.~Yuge, T.~Sagawa, A.~Sugita, H.~Hayakawa, Geometrical pumping in quantum
  transport: Quantum master equation approach, Physical Review B 86~(23) (2012)
  235308.

\bibitem{lu2022geometric}
J.~Lu, Z.~Wang, J.~Peng, C.~Wang, J.-H. Jiang, J.~Ren, Geometric thermodynamic
  uncertainty relation in a periodically driven thermoelectric heat engine,
  Physical Review B 105~(11) (2022) 115428.

\bibitem{walls1983squeezed}
D.~F. Walls, Squeezed states of light, nature 306~(5939) (1983) 141--146.

\bibitem{schnabel2017squeezed}
R.~Schnabel, Squeezed states of light and their applications in laser
  interferometers, Physics Reports 684 (2017) 1--51.

\bibitem{dodonov1991physical}
V.~Dodonov, A.~Klimov, V.~Man'ko, Physical significance of correlated and
  squeezed states, in: Group Theoretical Methods in Physics, Springer, 1991,
  pp. 450--456.

\bibitem{puri1997coherent}
R.~Puri, Coherent and squeezed states on physical basis, pramana 48~(3) (1997)
  787--797.

\bibitem{dupays2021shortcuts}
L.~Dupays, A.~Chenu, Shortcuts to squeezed thermal states, Quantum 5 (2021)
  449.

\bibitem{abebe2021interaction}
T.~Abebe, D.~Jobir, C.~Gashu, E.~Mosisa, Interaction of two-level atom with
  squeezed vacuum reservoir, Advances in Mathematical Physics 2021 (2021).

\bibitem{holmes2019coherent}
Z.~Holmes, S.~Weidt, D.~Jennings, J.~Anders, F.~Mintert, Coherent fluctuation
  relations: from the abstract to the concrete, Quantum 3 (2019) 124.

\bibitem{yadalam2022counting}
H.~K. Yadalam, B.~K. Agarwalla, U.~Harbola, Counting statistics of energy
  transport across squeezed thermal reservoirs, arXiv preprint arXiv:2202.04011
  (2022).

\bibitem{manzano2016entropy}
G.~Manzano, F.~Galve, R.~Zambrini, J.~M. Parrondo, Entropy production and
  thermodynamic power of the squeezed thermal reservoir, Physical Review E
  93~(5) (2016) 052120.

\bibitem{niedenzu2016operation}
W.~Niedenzu, D.~Gelbwaser-Klimovsky, A.~G. Kofman, G.~Kurizki, On the operation
  of machines powered by quantum non-thermal baths, New Journal of Physics
  18~(8) (2016) 083012.

\bibitem{agarwalla2017quantum}
B.~K. Agarwalla, J.-H. Jiang, D.~Segal, Quantum efficiency bound for continuous
  heat engines coupled to noncanonical reservoirs, Physical Review B 96~(10)
  (2017) 104304.

\bibitem{klaers2017squeezed}
J.~Klaers, S.~Faelt, A.~Imamoglu, E.~Togan, Squeezed thermal reservoirs as a
  resource for a nanomechanical engine beyond the carnot limit, Physical Review
  X 7~(3) (2017) 031044.

\bibitem{newman2017performance}
D.~Newman, F.~Mintert, A.~Nazir, Performance of a quantum heat engine at strong
  reservoir coupling, Physical Review E 95~(3) (2017) 032139.

\bibitem{ourjoumtsev2011observation}
A.~Ourjoumtsev, A.~Kubanek, M.~Koch, C.~Sames, P.~W. Pinkse, G.~Rempe, K.~Murr,
  Observation of squeezed light from one atom excited with two photons, Nature
  474~(7353) (2011) 623--626.

\bibitem{mehmet2010observation}
M.~Mehmet, H.~Vahlbruch, N.~Lastzka, K.~Danzmann, R.~Schnabel, Observation of
  squeezed states with strong photon-number oscillations, Physical Review A
  81~(1) (2010) 013814.

\bibitem{huang2012effects}
X.~Huang, T.~Wang, X.~Yi, et~al., Effects of reservoir squeezing on quantum
  systems and work extraction, Physical Review E 86~(5) (2012) 051105.

\bibitem{hsiang2021fluctuation}
J.-T. Hsiang, B.-L. Hu, Fluctuation--dissipation relation for a quantum
  brownian oscillator in a parametrically squeezed thermal field, Annals of
  Physics 433 (2021) 168594.

\bibitem{talkner2013statistics}
P.~Talkner, M.~Morillo, J.~Yi, P.~H{\"a}nggi, Statistics of work and
  fluctuation theorems for microcanonical initial states, New Journal of
  Physics 15~(9) (2013) 095001.

\bibitem{wang2021nonequilibrium}
C.~Wang, H.~Chen, J.-Q. Liao, Nonequilibrium thermal transport and photon
  squeezing in a quadratic qubit-resonator system, Physical Review A 104~(3)
  (2021) 033701.

\bibitem{giraldi2014coherence}
F.~Giraldi, F.~Petruccione, Coherence in a dissipative two-level system, The
  European Physical Journal D 68~(6) (2014) 1--7.

\bibitem{pekola2021colloquium}
J.~P. Pekola, B.~Karimi, Colloquium: Quantum heat transport in condensed matter
  systems, Reviews of Modern Physics 93~(4) (2021) 041001.

\bibitem{esposito2009nonequilibrium}
M.~Esposito, U.~Harbola, S.~Mukamel, Nonequilibrium fluctuations, fluctuation
  theorems, and counting statistics in quantum systems, Reviews of modern
  physics 81~(4) (2009) 1665.

\bibitem{PhysRevE.73.026109}
D.~Segal, A.~Nitzan,
  \href{https://link.aps.org/doi/10.1103/PhysRevE.73.026109}{Molecular heat
  pump}, Phys. Rev. E 73 (2006) 026109.
\newblock \href {https://doi.org/10.1103/PhysRevE.73.026109}
  {\path{doi:10.1103/PhysRevE.73.026109}}.
\newline\urlprefix\url{https://link.aps.org/doi/10.1103/PhysRevE.73.026109}

\bibitem{carpio2021quantum}
P.~Carpio-Martinez, G.~Hanna, Quantum bath effects on nonequilibrium heat
  transport in model molecular junctions, The Journal of Chemical Physics
  154~(9) (2021).

\bibitem{kowalewska2001generalized}
A.~Kowalewska-Kudlaszyk, R.~Tana{\'s}, Generalized master equation for a
  two-level atom in a strong field and tailored reservoirs, journal of modern
  optics 48~(2) (2001) 347--370.

\bibitem{saryal2019thermodynamic}
S.~Saryal, H.~M. Friedman, D.~Segal, B.~K. Agarwalla, Thermodynamic uncertainty
  relation in thermal transport, Physical Review E 100~(4) (2019) 042101.

\bibitem{lu2021nonequilibrium}
Y.~Lu, N.~Lambert, A.~F. Kockum, K.~Funo, A.~Bengtsson, S.~Gasparinetti,
  F.~Nori, P.~Delsing, Steady-state heat transport and work with a single
  artificial atom coupled to a waveguide: Emission without external driving,
  PRX Quantum 3~(2) (2022) 020305.

\bibitem{harbola2007statistics}
U.~Harbola, M.~Esposito, S.~Mukamel, Statistics and fluctuation theorem for
  boson and fermion transport through mesoscopic junctions, Physical Review B
  76~(8) (2007) 085408.

\bibitem{li2017production}
S.-W. Li, et~al., Production rate of the system-bath mutual information,
  Physical Review E 96~(1) (2017) 012139.

\bibitem{tanas2002squeezing}
R.~Tanas, Squeezing and squeezing-like terms in the master equation for a
  two-level atom in strong fields, Journal of Optics B: Quantum and
  Semiclassical Optics 4~(3) (2002) S142.

\bibitem{sarmah2023nonequilibrium}
M.~J. Sarmah, A.~Bansal, H.~P. Goswami, Nonequilibrium fluctuations in boson
  transport through squeezed reservoirs, Physica A: Statistical Mechanics and
  its Applications 615 (2023) 128620.

\bibitem{pietzonka2017finite}
P.~Pietzonka, F.~Ritort, U.~Seifert, Finite-time generalization of the
  thermodynamic uncertainty relation, Physical Review E 96~(1) (2017) 012101.

\bibitem{yuge2013geometrical}
T.~Yuge, T.~Sagawa, A.~Sugita, H.~Hayakawa, Geometrical excess entropy
  production in nonequilibrium quantum systems, Journal of Statistical Physics
  153 (2013) 412--441.

\bibitem{dodonov2002nonclassical}
V.~Dodonov, Nonclassical'states in quantum optics: asqueezed'review of the
  first 75 years, Journal of Optics B: Quantum and Semiclassical Optics 4~(1)
  (2002) R1.

\bibitem{bagrets2003full}
D.~Bagrets, Y.~V. Nazarov, Full counting statistics of charge transfer in
  coulomb blockade systems, Physical Review B 67~(8) (2003) 085316.

\end{thebibliography}






\end{document}